\documentclass{iau}

\usepackage{amsmath}
\usepackage{graphicx}
\usepackage{multirow}

\begin{document}

\lefttitle{Sepideh Ghaziasgar et al.}
\righttitle{Photometric vs. Spectroscopic Labels of Dusty Stellar Sources classification}

\jnlPage{1}{7}
\jnlDoiYr{2025}
\doival{10.1017/xxxxx}

\aopheadtitle{Proceedings IAU Symposium}
\editors{C. Sterken,  J. Hearnshaw \&  D. Valls-Gabaud, eds.}

\title{Comparison of Photometric and Spectroscopic Labels in Classifying Dusty Stellar Sources Using Machine Learning in the Magellanic Clouds}

\author{
Sepideh Ghaziasgar$^{1}$\textsuperscript{*},
Mahdi Abdollahi$^{1}$,
Atefeh Javadi$^{1}$,
Jacco Th. van Loon$^{2}$,
Iain McDonald$^{3}$,
Joana Oliveira$^{2}$,
Habib G. Khosroshahi$^{1,5}$
}

\affiliation{
$^{1}$School of Astronomy, Institute for Research in Fundamental Sciences (IPM), P.O. Box 195683-6613, Tehran, Iran; \texttt{sepideh.ghaziasgar@ipm.ir}\\[0.2cm]
$^{2}$Lennard-Jones Laboratories, Keele University, Keele, ST5 5BG, UK \\[0.2cm]
$^{3}$Jodrell Bank Centre for Astrophysics, Alan Turing Building, University of Manchester, Manchester, M13 9PL, UK \\[0.2cm]
$^{5}$Iranian National Observatory, Institute for Research in Fundamental Sciences (IPM), Tehran, Iran\\[0.2cm]
}

\begin{abstract}
Dusty stellar sources, including young stellar objects (YSOs) and evolved stars such as oxygen- and carbon-rich AGBs (OAGBs, CAGBs), red supergiants (RSGs), and post-AGB stars (PAGBs), play a key role in the chemical enrichment of galaxies. Photometric surveys in the Magellanic Clouds have cataloged many such objects, but their classifications are often uncertain due to overlaps between populations. We trained machine learning models on spectroscopically labeled data from the SAGE project and applied them to photometric catalogs. The spectroscopic model achieves about 89\% accuracy. Applied to photometric labels, nearly all OAGBs are correctly identified, and YSOs have a 95\% confirmation rate. In contrast, 16\% of CAGBs are reclassified as OAGBs, only 8\% of RSGs retain their labels, and fewer than half of PAGBs are confirmed. Photometry is thus reliable for abundant populations with distinct signatures, but spectroscopic confirmation remains essential for rare or overlapping stellar classes.
\end{abstract}


\maketitle
\section{Introduction}

Dusty stellar sources play a fundamental role in the life cycle of galaxies by returning heavy elements to the interstellar medium. 
These objects include young stellar objects (YSOs), which trace ongoing star formation, as well as evolved stars such as asymptotic giant branch (AGB) stars, red supergiants (RSGs), and post-AGB stars (PAGBs) \citep{2023ApJ...948...63A,Boyer2011b,2025arXiv250716766B,2025arXiv250805596M,2018A&ARv..26....1H,2011MNRAS.411..263J,2013MNRAS.432.2824J,2015MNRAS.447.3973J}.

Because of the high luminosities and red colors of dusty stellar sources, these populations are particularly well traced in the Magellanic Clouds (MCs), two nearby dwarf galaxies with low metallicities \citep{1992ApJ-metalicity-russel,LMC-SMC-2009A&A...496..399S,Pietrzy-lmc-2013Natur.495...76P,Subramanian-2011ASInC...3..144S}, making them ideal laboratories for studying dusty stellar evolution.

Traditionally, the identification of such objects has relied on photometric surveys using color–magnitude diagrams (CMDs) and infrared excesses. 
While photometry enables the construction of large catalogs at relatively low observational cost, it suffers from ambiguities: distinct stellar classes can occupy overlapping regions in photometric space.  
In contrast, spectroscopy provides a more secure diagnostic by revealing the chemical and physical properties of stellar atmospheres and circumstellar dust, but spectroscopic surveys are much more limited in size.

With the growing availability of both photometric and spectroscopic datasets from surveys such as the \textit{Spitzer} SAGE and SAGE-Spec programs \citep{Kemper2010,Woods2011,Ruffle2015,Jones2017}, it has become possible to directly assess the reliability of photometric classifications by comparing them with spectroscopic benchmarks.

Unsupervised machine learning has also been applied in this context; for example, \citet{2022MNRAS.515.6046P} used t-distributed Stochastic Neighbour Embedding (t-SNE) to identify sources dominated by AGN-dust emission in regions of the sky that include the Magellanic Clouds. In contrast, our study relies exclusively on supervised learning, training models on spectroscopically confirmed sources and then applying them to large photometric catalogs.

Machine learning provides a powerful means of conducting such comparisons \citep{2019arXiv190407248B}. In this study, we used supervised models trained on spectroscopic labels of dusty stellar populations in the MCs \citep{2024BAO-Ghaziasgar,2025IAUS..368...67G,2025ApJ-Ghaziasgar}. The tested algorithms, as described in \citet{2025ApJ-Ghaziasgar}, included Probabilistic Random Forest (PRF), Random Forest (RF), K-Nearest Neighbors (KNN), C-Support Vector Classification (SVC; polynomial and RBF kernels), and Gaussian Naive Bayes (GNB). Among these, PRF \citep{2019AJ....157...16R,2021MNRAS.507.5106K,2022MNRAS.517..140K,2025MNRAS.537.1028P} provided the best performance, as it naturally accounts for uncertainties in both features and labels. By comparing the labels predicted by the model trained on spectroscopic labels and photometric labels, we aimed to quantify both the strengths and the limitations of photometric classification, especially for less numerous or overlapping stellar classes.

\section{Data and Method}

The spectroscopic dataset originates from the SAGE-Spec program and archival \textit{Spitzer}/IRS observations, encompassing more than 600 dusty stellar sources in the LMC and SMC \citep{Kemper2010,Woods2011,Ruffle2015,Jones2017}. These sources were spectroscopically labeled into five classes: YSO, CAGB, OAGB, RSG, and PAGB. 

For the photometric sample, we compiled catalogs from major surveys of the LMC and SMC \citep{Whitney2008, Srinivasan2009, 2009ApJS..184..172G, Riebel2010, Boyer2011b, Kamath2014, Kamath2015, Yang2018, Yang2019, Yang2020, Yang2021}. Photometric classifications are generally based on positions in CMDs and infrared excess, but they may suffer from ambiguities where stellar classes overlap. We cross-matched the photometric catalogs with the spectroscopic set, removed duplicates, and applied the same preprocessing steps (feature selection, imputation, extinction, and distance corrections) as described in \citet{2025ApJ-Ghaziasgar}.\

To assess the reliability of the photometric classifications reported in those mentioned above, we apply models trained on spectroscopic labels presented by \citet{2024BAO-Ghaziasgar,2025ApJ-Ghaziasgar}. As discussed earlier, several models were tested and validated, and the probabilistic random forest (PRF) model was ultimately identified as the best performer.\

\section{Comparison Results}

\begin{figure}[htbp]

    \centering
    \includegraphics[width=0.7\linewidth]{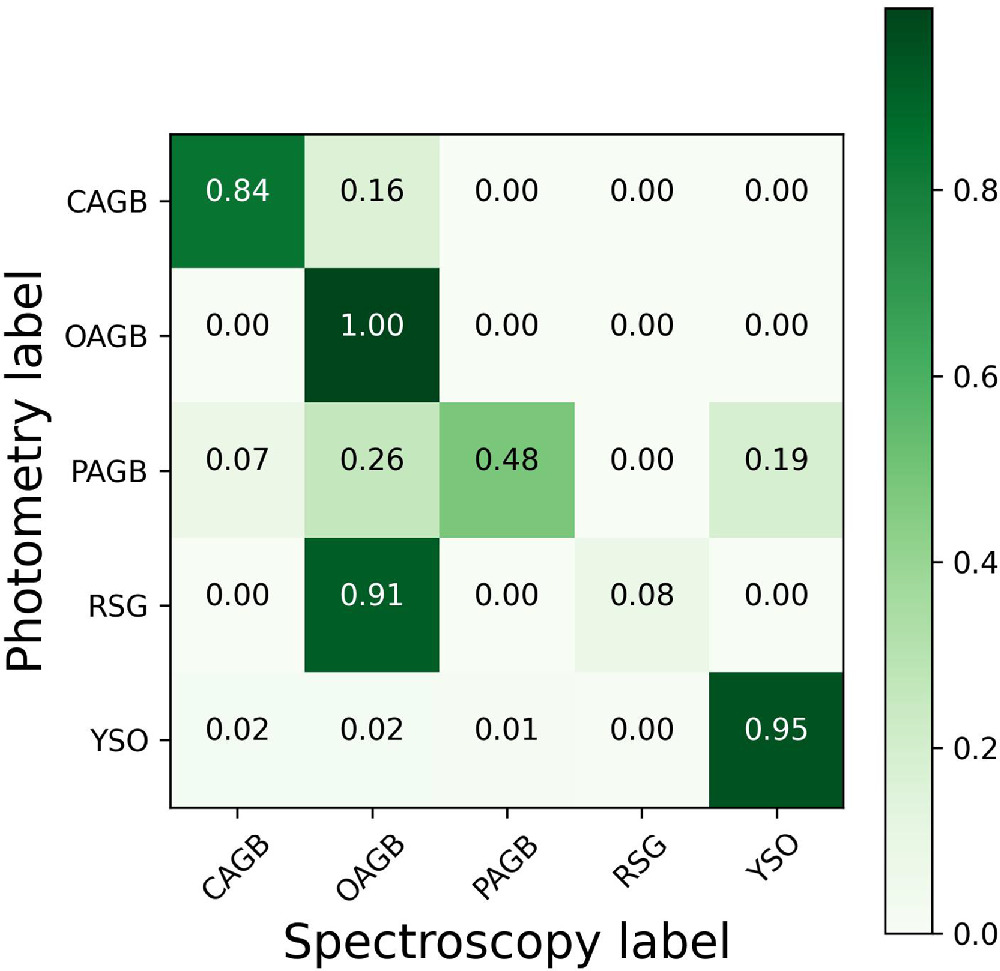}
    \caption{Confusion matrix comparing photometric and spectroscopic labels.}
    \label{fig:comp-matrix}

\end{figure}

The comparison between the predicted data and the photometric labels reveals that oxygen-rich AGB stars are almost perfectly recovered, and young stellar objects show a high confirmation rate of about 95\%. In contrast, roughly 16\% of carbon-rich AGB stars are reassigned as OAGBs, reflecting overlaps in CMD space. Red supergiants are particularly problematic: only about 8\% of photometric RSG labels are confirmed, with most misclassified as OAGBs. Post-AGB stars show the highest variability, with fewer than half confirmed.  
A more detailed view of these misclassifications, based on confusion matrices for each class, is presented in Figure~\ref{fig:comp-matrix}.\ 

Overall, photometric classification is reliable for abundant populations with distinctive infrared signatures, such as OAGBs and YSOs; however, it is less effective for rare or overlapping stellar classes. These results underscore the importance of spectroscopic confirmation in such cases, as well as the value of spectroscopically trained machine learning models in refining large photometric catalogs.\
\begin{table}[ht!]
\centering
\caption{Photometric classification of dusty stars in the Magellanic Clouds after cross-matching all catalogs and removing duplicates.}\

\label{tab:photometric-stats}

\resizebox{0.4\linewidth}{!}{ 
\begin{tabular}{l r}
\hline
\textbf{Stellar Type} & \textbf{Number} \\
\hline
CAGB  & 8,537  \\
OAGB  & 37,950 \\
PAGB  & 45     \\
RSG   & 6,132  \\
YSO   & 1,642  \\
\hline
\textbf{Total} & \textbf{54,306} \\
\hline
\end{tabular}
}
\end{table}

\end{document}